\documentstyle[amssymb,preprint,aps]{revtex}
\begin{document}
\title{Global aspects of gravitomagnetism}
\author{A. Barros$^{1}$ , V. B. Bezerra$^{2}$ and C. Romero$^{2}$}
\address{$^{1}$Departamento de F\'{i}sica, Universidade Federal de Roraima,\\
69310-270, Boa Vista, RR, Brazil\\
$^{2}$Departamento de F\'{i}sica, Universidade Federal da Para\'{i}ba, Caixa%
\\
Postal 5008, 58051-970, Jo\~{a}o Pessoa, PB, Brazil\\
e-mail: cromero@fisica.ufpb.br}
\maketitle

\begin{abstract}
We consider global properties of gravitomagnetism by investigating the
gravitomagnetic field of a rotating cosmic string. We show that although the
gravitomagnetic field produced by such a configuration of matter vanishes
locally, it can be detected globally. In this context we discuss the
gravitational analogue of the Aharonov-Bohm effect. 


\end{abstract}

\section{ Introduction}

\bigskip The conjecture that mass currents should generate a field called,
by analogy with eletromagnetism, the gravitomagnetic field, goes back to the
beginnings of general relativity\cite{Lense}. Indeed, according to general
relativity, moving or rotating matter should produce a contribution to the
gravitational field that is the analogue of the magnetic field of a moving
charge or magnetic dipole. This field would be expected to manifest itself
in a number of effects, such as the Lense-Thirring precession\cite{Wheeler},
the gravitomagnetic time delay\cite{ciufolini}, change in the phase of
electromagnetic waves\cite{Cohen}, among others.

Recently, interest in the subject has been boosted by the concrete
possibility that gravitomagnetic effects might be measured with the current
technology of laser ranged satellites (LAGEOS and LAGEOS II)\cite{lageos}.
It is important to mention the Relativity Gyroscope Experiment\ (Gravity
Probe B), a space mission to be launched this year whose aim is to detect
gravitomagnetism effects directly. It is expected that these experimental
programs will open new possibilities of testing general relativity and other
metric theories of gravity \cite{Camacho,Will}.

\ In the usual approach to gravitomagnetism one generally assumes the weak
field and slow motion approximation of general relativity. It is in this
limit that gravity parallels electromagnetism. On the other hand,
interesting attempts to give a precise characterization of gravitomagnetism,
which does not depend on the linearization of general relativity, have been
made. For instance, an invariant characterization of gravitomagnetism (valid
for any metric theory) has been proposed, which uses the pseudoscalar $%
^{\ast }RR\equiv \frac{1}{2}\epsilon ^{\alpha \beta \sigma \rho }R_{\sigma
\rho }^{....\mu \nu }R_{\alpha \beta \mu \nu }$ as a measure of the presence
of gravitomagnetism in a given spacetime\cite{curvatura}. However, since
curvature is a local concept, this characterization would not work if the
gravitational field manifest itself only globally, as in the case of
spacetimes generated by cosmic strings. Therefore it seems to be of interest
to develop a way of tackling such special cases. Clearly, a technique which
seems to be quite appropriate to investigate global aspects of
gravitomagnetism is holonomy theory, and, in fact, it has been employed
recently to investigate the holonomic manifestation of the gravitomagnetic
clock effect\cite{maarteens}.

This paper is organized as follows. In Section II, we introduce the reader
to the basic ideas of local gravitomagnetism. In Section III, we consider
the spacetime generated by a spinning cosmic string and then investigate the
gravitomagnetism associated with this kind of matter configuration. The
Aharonov-Bohm effect in connection with the gravitomagnetism will be
examined in Section IV. Finally, Section V is devoted to our conclusion and
some remarks.

\section{The gravitomagnetic field in general relativity \ \ }

Let us recall that in the weak field approximation of general relativity we
assume that the metric tensor $g_{\mu \nu }$ deviates only slightly from the
flat spacetime metric tensor. In other words, we assume that $g_{\mu \nu
}=\eta _{\mu \nu }+h_{\mu \nu }$, where $\eta _{\mu \nu }=diag(-1,1,1,1)$
denotes Minkowski metric tensor and $h_{\mu \nu }$ is a small perturbation
term. Then, by keeping only first-order \ terms in $h_{\mu \nu }$ and
adopting the usual harmonic coordinate gauge $\left( h_{\upsilon }^{\mu }-%
\frac{1}{2}\delta _{\nu }^{\mu }h\right) ,_{\mu }=0$, the Einstein equations
become

\begin{equation}
\square \overline{h}_{\mu \nu }=-\frac{16\pi G}{c^{4}}T_{\mu \nu }
\label{weakfieldgr}
\end{equation}
where $\overline{h}_{\nu }^{\mu }=h_{\upsilon }^{\mu }-\frac{1}{2}\delta
_{\nu }^{\mu }h$ and $h$ denotes the trace of $h_{\upsilon }^{\mu }$\cite
{Landau}.

We now assume a perfect fluid matter configuration and slow motion. If $%
\sigma $ denotes the mass density and $v_{i\text{ }}$the velocity
components, then (\ref{weakfieldgr}) yields 
\begin{equation}
\square \overline{h}_{00}=-\frac{16\pi G}{c^{2}}\sigma
\label{gravitoelectric}
\end{equation}
\begin{equation}
\square \overline{h}_{0i}=\frac{16\pi G}{c^{3}}\sigma v_{i}
\label{gravitomagnetic}
\end{equation}
where terms such as $p$ and $v_{i}v_{j}/c^{4}$ have been neglected. Let us
now specialize the equations above to the case of a stationary gravitational
field of a slowly rotating body. Then, far from the source we have 
\begin{equation}
\nabla ^{2}\left( \frac{c^{2}\overline{h}_{00}}{4}\right) \equiv \nabla
^{2}(\Phi _{g})=-4\pi G\sigma  \label{ro}
\end{equation}
\begin{equation}
\nabla ^{2}\overline{h}_{0i}=\frac{16\pi G}{c^{3}}\sigma v_{i}  \label{ui}
\end{equation}
from which it follows that 
\begin{equation}
\Phi _{g}=\frac{GM}{r}  \label{fi}
\end{equation}
\begin{equation}
\overrightarrow{\overline{h}}=-\frac{2G(\overrightarrow{J}\times 
\overrightarrow{r})}{c^{3}r^{3}}\equiv -\frac{2\overrightarrow{A}_{g}}{c^{2}}
\label{ag}
\end{equation}
where $\overline{h}_{0i}$ are the components of the vector $\overrightarrow{%
\overline{h}}$, $M$ and $\overrightarrow{J}$ are the total mass and angular
momentum of the source, respectively. In close analogy with electrodynamics
we define the gravitoelectric field to be $\overrightarrow{E_{g}}=-{\bf %
\nabla }\Phi _{g}$ and the gravitomagnetic field to be $\overrightarrow{B}%
_{g}=\overrightarrow{{\bf \nabla }}\times \overrightarrow{A}_{g}$. It is
interesting to see that the condition $\overline{h}^{\mu \nu },_{\mu }=0$
leads to ${\bf \nabla }\cdot \overrightarrow{A}_{g}=0$ (analogous to the
Coulomb gauge of electromagnetism).

Let us note that for the case of a slowly rotating sphere with angular
momentum $\overrightarrow{J}=(0,0,J)$, we obtain from (\ref{ag}) in
spherical coordinates 
\begin{equation}
\overline{h}_{0\varphi }=h_{0\varphi }=-\frac{2JG}{rc^{3}}\sin ^{2}\theta
\label{kerr}
\end{equation}
Recalling that the Kerr metric in Boyer-Lindquist coordinates in the weak
field and slow motion limit is given by\cite{Boyer} 
\begin{equation}
ds^{2}=-\left( 1-\frac{2MG}{rc^{2}}\right) c^{2}dt^{2}+\left( 1+\frac{2MG}{%
rc^{2}}\right) dr^{2}+r^{2}(d\theta ^{2}+\sin ^{2}\theta d\varphi ^{2})-%
\frac{4JG}{rc^{3}}\sin ^{2}\theta cdtd\varphi  \label{Kerrmetric}
\end{equation}
we see that $\overline{h}_{0\varphi }$ is the $g_{0\varphi }$ component of (%
\ref{Kerrmetric}).

It is worth noting that one can easily show by using the geodesic equation 
\begin{equation}
\frac{d^{2}x^{\mu }}{ds^{2}}+\Gamma _{\alpha \beta }^{\mu }\frac{dx^{\alpha }%
}{ds}\frac{dx^{\beta }}{ds}=0  \label{geodesic}
\end{equation}
in the slow motion and weak field approximation, that \cite{mashh} 
\begin{equation}
\frac{d^{2}\overrightarrow{r}}{dt^{2}}\cong \left( \overrightarrow{E_{g}}+%
\frac{2}{c}\frac{d\overrightarrow{r}}{dt}\times \overrightarrow{B}_{g}\right)
\label{Lorentz}
\end{equation}
From the above we see that both the gravitoelectric and the gravitomagnetic
field are essentially local physical entities. It turns out, however, that
nonlocal properties of gravitomagnetism may appear, for example, when we
consider the spacetime generated by a spinning cosmic string. Let us
consider this case in the next section.

\section{\protect\bigskip The gravitomagnetism of a spinning cosmic string}

As objects whose existence in the early Universe has been predicted by GUT
theories, cosmic strings have attracted a lot of interest in the last two
decades\cite{Vilenkin}. It was shown by Vilenkin that, according to general
relativity, the spacetime generated by a static cosmic string has a conical
singularity\cite{Vilenkin2,Gott,Hiscock}. It is this conicity that gives
rise to a series of notable gravitational effects, mainly in connection with
the fact that the geometry is locally flat albeit it possesses nontrivial
global properties. Solutions corresponding to spinning cosmic strings were
also found\cite{Mazur,Jensen,Lemos,Jensen2}. It has been shown that, in
addition to the conical topology of the static string, the geometry of a
spinning cosmic string has a helical time structure which allows for the
existence of closed timelike curves near the source\cite{Novello}, as well
as gravitational time delay effects\cite{Harari}. On the other hand, the
gravitational time delay in ray propagation due to rotating bodies has been
shown recently to be an effect directly related to the gravitomagnetic field
produced by the bodies\cite{ciufolini}, which is rather suggestive of a
possible presence of gravitomagnetic effects caused by a spinning cosmic
string.

The spacetime of a spinning cosmic string is described by the metric\cite
{Mazur} 
\begin{equation}
ds^{2}=-(cdt+\frac{4GJ}{c^{3}}d\phi )^{2}+d\rho ^{2}+a^{2}\rho ^{2}d\phi
^{2}+dz^{2}  \label{spinning}
\end{equation}
where $a=1-\frac{4G\mu }{c^{2}}$ \ is a measure of the angle deficit and $J$
denotes the angular momentum per unit of length of the string. This solution
was obtained by a lifting from the $(2+1)$-dimensional spinning particle
solution \cite{Deser} to $(3+1)$ dimensions. This was performed by simply
adding $dz%
%TCIMACRO{\UNICODE[m]{0xb2}}%
%BeginExpansion
{{}^2}%
%EndExpansion
$ to the metric of the spinning particle spacetime. It is worth noting that
the spacetime of a spinning cosmic string is locally flat everywhere except
at the source. This is immediately seen by carrying out the coordinate
transformation 
\[
t^{\prime }=t+4\frac{GJ}{c^{4}}\phi ,\text{ \ }\phi ^{\prime }=a\phi ,\text{
\ }\rho ^{\prime }=\rho ,\text{ }z^{\prime }=z 
\]
Note that if $J=0$, (\ref{spinning}) represents the metric $\gamma _{\mu \nu
}$ of the spacetime generated by a static vacuum cosmic string\cite
{Vilenkin2,Gott,Hiscock}. If we consider low rates of spin motion, then the
metric (\ref{spinning}) may be written as 
\begin{equation}
ds^{2}=-c^{2}dt^{2}+d\rho ^{2}+a^{2}\rho ^{2}d\phi ^{2}+dz^{2}-\frac{8GJ}{%
c^{2}}dtd\phi  \label{perturbation}
\end{equation}
Thus, we may view the line element above as a perturbation of the metric of
the static cosmic string spacetime, i.e. it has the form $ds^{2}=g_{\mu \nu
}dx^{\mu }dx^{\nu }$, with $g_{\mu \nu }=\gamma _{\mu \nu }+h_{\mu \upsilon
} $ , where $h_{\mu \upsilon }$ is to be regarded as a small perturbation of 
$\gamma _{\mu \nu }$. Thus, in the same way as one can introduce the
gravitomagnetic field directly from the Kerr metric in the weak field and
slow motion approximation\cite{Lense}, it seems natural to define the
gravitomagnetic vector potential by 
\begin{equation}
\overrightarrow{A}_{g}=-\frac{c^{2}}{2}\overrightarrow{h}=-\frac{c^{2}}{2}%
(h_{0\rho ,}h_{0\phi ,}h_{0z})=(0,\frac{4GJ}{c},0)  \label{potential}
\end{equation}
where $\overrightarrow{h}\equiv (h_{01},h_{02},h_{03})$.

Therefore, from the above equation, we conclude that $\overrightarrow{B}_{g}=%
\overrightarrow{{\bf \nabla }}\times \overrightarrow{A}_{g}=0$, i.e. the
gravitomagnetic field vanishes locally. However, taking into account the
nontrivial topology of the spacetime (\ref{spinning}) one would wonder
whether this quantity might somehow be detected globally. In the next
section we shall show that, in fact, a global property of the
gravitomagnetism associated with the spinning cosmic string is what is
called the gravitational analogue of Aharonov-Bohm effect\cite{Aharonov}.

\section{A gravitational analogue of the Aharonov-Bohm effect}

Gravitational analogues of the Aharonov-Bohm effect have been studied by
Dowker\cite{Dowker} and others \cite{Ford,Valdir}, mainly in connection with
the conical spacetime\cite{Marder}. The essential requirement for the
analogy is the existence of a spacetime whose curvature vanishes everywhere
except in some localized region. In such spacetimes it is possible that
particles travelling along paths where the curvature vanishes still feel a
physical effect in much the same way as the wave function of a charged
particle undergoes a change of phase 
\begin{equation}
\delta \alpha =\frac{e}{\hbar c}\oint_{C}\overrightarrow{A}\cdot d%
\overrightarrow{l}  \label{phase}
\end{equation}
when the integration above is carried out around a closed circuit $C$
including a region where the magnetic field $\overrightarrow{B\text{ }}$ is
nonvanishing, even if the wave function is zero in this region, an effect
which has no counterpart in classical physics and has been measured in a
number of experiments\cite{Tonomura}.

Other effects analogous to the electromagnetic Aharonov-Bohm effect exist in
a classical context, such as the Sagnac effect\cite{Post} in general
relativity, which consists in a phase shift between two beams of light
traversing in opposite directions the same path around a rotating mass
distribution.

Now, let us turn our attention to the change of phase induced by the
gravitomagnetic vector potential given by (\ref{potential}). Clearly, it is
given by

\begin{equation}
\delta \alpha _{g}=\kappa \oint_{C}\overrightarrow{A_{g}}\cdot d%
\overrightarrow{l}=\frac{8\pi \kappa GJ}{c}  \label{phase}
\end{equation}
where we have introduced the constant $\kappa $ in order to ensure that $%
\delta \alpha _{g}$ is dimensionless.

Therefore, we are led to the conclusion that although locally we have no
gravitomagnetism, yet there is still a global effect, namely, a change on
the phase, due to the gravitomagnetic potential, hence in close analogy with
the Aharonov-Bohm effect. Incidentally, this result coincides with the
holonomy\cite{Valdir} for general curves in the $xy-plane$, which enclose
the spinning cosmic string, when $a=1$. A similar result can also be
obtained from the Schr\"{o}dinger equation governing the motion of a
particle with electric charge $e$, moving in the field of an infinitely thin
solenoid which carries a magnetic flux $\Phi $, when we replace $\frac{e\Phi 
}{2\pi }$ by $GJ$. These results seems to suggest that the choice we have
made for $\overrightarrow{A_{g}}$, given by (\ref{potential}), in the
framework of the analogy between electromagnetism and gravitation, is
correct.

\section{\protect\bigskip Final remarks}

Another global gravitomagnetic effect, which happens in the absence of the
local gravitomagnetic field and is due to the spacetime topology only, is
the time delay between the arrival time of two particles that describe
arbitrary closed paths in opposite directions around a spinning cosmic
string. It has been shown that this time delay, a gravitomagnetic observable
effect at a classical level \cite{Harari}, is given by $\Delta \tau =\frac{%
16\pi GJ}{c^{4}}=\frac{4\pi }{c^{4}}A_{g}$.

Gravitomagnetism is often regarded as a manifestation of the way inertia is
incorporated into general relativity. According to this view, the
measurement of gravitomagnetic effects would certainly be considered an
experimental support of what is called the weak general relativistic
interpretation of the Mach principle \cite{Lense}. However, this
interpretation is concerned with {\it local }inertial forces. The fact that
gravitomagnetism may manifest itself also {\it globally} would suggest one
to conjecture that the Mach principle may also possess a global character.

\end{document}